\documentstyle[aps,preprint]{revtex}
\begin{document}
\draft
\title { $^{35}$Cl+$^{12}$C ASYMMETRICAL FISSION EXCITATION FUNCTIONS }

\author { C. Beck, D. Mahboub, R. Nouicer, T. Matsuse \cite{adr1}, B. Djerroud
\cite{adr2}, R.M. Freeman, F. Haas, \\
A. Hachem \cite{adr3}, A. Morsad \cite{adr4} and M. Youlal }

\address{\it Centre de Recherches Nucl\'eaires, Institut National de Physique
Nucl\'eaire et de Physique des Particules - Centre National de la Recherche
Scientifique/Universit\'e Louis Pasteur,B.P.28, F-67037 Strasbourg Cedex 2,
France }

\author { S.J. Sanders }

\address {\it The University of Kansas, Department of Physics and Astronomy,
Lawrence KS 66045, USA }

\author { R. Dayras, J.P. Wieleczko \cite{adr5}, E. Berthoumieux, R. Legrain,
E. Pollacco }

\address { DAPNIA/SPhN, C.E. Saclay, F-91191 Gif sur Yvette Cedex, France }

\author { Sl. Cavallaro, E. De Filippo, G. Lanzan\`o, A. Pagano and
M.L. Sperduto }

\address {\it Dipartimento di Fisica dell'Universit\'a di Catania, INFN and
LNS Catania, I-95129 Catania, Italy }

\date{\today}
\maketitle

\newpage

\begin{abstract}
{ The fully energy-damped yields from the $^{35}$Cl+$^{12}$C reaction have been
systematically investigated using particle-particle coincidence techniques at a
$^{35}$Cl bombarding energy of $\sim$ 8 MeV/nucleon. The fragment-fragment
correlation data show that the majority of events arises from a binary-decay
process with rather large numbers of secondary light-charged particles emitted
from the two excited exit fragments. No evidence is observed for ternary
break-up events. The binary-process results of the present measurement, along
with those of earlier, inclusive experimental data obtained at several lower
bombarding energies are compared with predictions of two different kinds of
statistical model calculations. These calculations are performed using the
transition-state formalism and the Extended Hauser-Feshbach method and are
based on the available phase space at the saddle point and scission point of
the compound nucleus, respectively. The methods give comparable predictions and
are both in good agreement with the experimental results thus confirming the
fusion-fission origin of the fully-damped yields. The similarity of the
predictions for the two models supports the claim that the scission point
configuration is very close to that of the saddle point for the light $^{47}$V
mass-asymmetry-dependent fission barriers needed in the transition-state
calculation. }
\end{abstract}




\centerline {\bf I. INTRODUCTION }

\vskip 1.5 cm

The $^{35}$Cl+$^{12}$C binary-reaction channels have been studied over the past
few years \cite{Be89,Be92,Dj92} at several relatively low $^{35}$Cl bombarding
energies (E$_{lab}$ $\leq$ 200 MeV), in the framework of a more general
investigation of the $^{47}$V composite system \cite{Dj92,Ra91,Be93}. It has
been established that entrance channel effects do not play a significant role in
the
binary-decay processes of the $^{47}$V nucleus as populated by the three
studied entrance channels ($^{35}$Cl+$^{12}$C \cite{Be89,Be92},
$^{31}$P+$^{16}$O \cite{Ra91} and $^{23}$Na+$^{24}$Mg \cite{Be93}) at
comparable excitation energies (E$_{CN}^{*}$ = 59-64 MeV) and angular momenta.
This demonstrates that in each case the compound nucleus (CN) is formed after a
complete equilibration of the mass asymmetry and the shape degrees of freedom.
Further, the properties of the binary-decay channels are in agreement with the
expectations of the transition-state model \cite{Sa91} and thus suggest a
fusion-fission (FF) origin. The occurrence of FF rather than orbiting in
certain systems, such as the $^{47}$V dinucleus, has been the subject of large
number of discussions \cite{Be92,Dj92,Ra91,Be93,Sa91,Ma91,Ma96,Ca95}.
These have led, in most cases, to the conclusion that the FF process has to be
taken into account when exploring the limitations of the complete fusion
process at large angular momenta and high excitation energies. This is in
accordance with the qualitative expectations of the number of open channels
model \cite{Be94}.

In this paper we report on further inclusive measurements on FF yields of
$^{35}$Cl+$^{12}$C at a bombarding energy E$_{lab}$($^{35}$Cl) = 280 MeV. This
significantly extends the known complete FF excitation function for this
reaction. In addition, fragment-fragment correlation data have been collected
to check the relative importance of mechanisms other than secondary particle
emission for the high excitation energy (E$_{CN}^{*}$ = 85 MeV) in the compound
system reached by this reaction. Light-fragment emission from the reaction
fragments can lead to a ``charge deficit" between the observed and
entrance-channel charge. The charge deficits in the present measurement are
extracted from the coincidence data in order to verify that they follow the
systematic trend that has been found with previously published results, thus
confirming the binary nature of the reaction. No evidence is seen for the onset
of ternary processes for incident energies lower than 10 MeV/nucleon.

It is known that the secondary and sequential light-charged-particle emission
from the fully accelerated binary-decay fragments increases with energy.
Nevertheless, the properties of the primary fission fragments can be deduced,
even at energies as high as 280 MeV, by using the coincident data. As found at
the lower energies studied \cite{Be89,Be92,Dj92} the hypothesis of CN formation
followed by a statistical decay is supported by the present data. Further, the
measured FF excitation function for each fragment atomic number is found to be
compatible with the choice of mass-asymmetric-dependent fission barriers used
in the transition-state model calculations \cite{Sa91}.

A brief description of the experimental techniques is given in the next
section. The experimental results of the single and coincidence measurements
are presented in Sect.III. These results are then discussed in Sect.IV within
the framework of two distinctly different FF pictures based upon whether the
saddle point \cite{Sa91} or the scission point \cite{Ma91,Ma96} determines the
fission decay rates.

\vskip 2.5 cm

\centerline {\bf II. EXPERIMENTAL PROCEDURES }

\vskip 1.5 cm

The investigation of the $^{35}$Cl+$^{12}$C reaction has been achieved at
the Strasbourg 18 MV MP Tandem and the Saclay Booster Tandem facilities by
means of kinematical coincidence techniques. Further details of the
experimental methods can be found elsewhere \cite{Be92,Dj92,Be93} in the
experimental descriptions for the lower energy measurements. The experiments at
the highest energy were performed with the 278-280 MeV pulsed beams, provided
by the Saclay post-accelerator, focused onto a 100 $\mu$g/cm$^{2}$ thick,
self-supporting $^{12}$C target mounted at the center of the 2 meter wide
scattering chamber ``chambre 2000". Some data which will be
presented in the discussion were also recorded with a 255 $\mu$g/cm$^{2}$ thick
$^{24}$Mg target for a study of the $^{35}$Cl+$^{24}$Mg reaction \cite{Ca95}.
During the course of the experiment at 280 MeV the reaction products were
detected, in a singles mode, either with four small size $\Delta$E-E telescopes
(located in the 10$^{o}$-35$^{o}$ angular range with a 2$^{o}$ step increment),
each consisting of a gas ionization chamber followed by a 500 $\mu$m thick
Si(SB) detector (IC), or with a large size, movable, Bragg Curve Spectrometer
(BCS) \cite{Mo84} covering the 2.5$^{o}$-12$^{o}$ angular range with 1$^{o}$
steps. The experimental set-up for the coincidence experiment at E$_{lab}$ =
278 MeV was very similar to that used for the inclusive experiment and
consisted of seven ionization-chamber telescopes in the reaction plane between
- 37$^{o}$ and + 85$^{o}$.

The energy calibrations of the BCS and IC's were obtained using elastically
scattered $^{35}$Cl projectiles from a 100 $\mu$g/cm$^{2}$ thick Au target and
from the C and Mg targets, combined with measurements of $\alpha$ sources and a
calibration pulser. On an event-by-event basis, corrections were applied for
energy loss in the target and IC's window foils and for the pulse height defect
in the Si detectors \cite{Ka74}. The absolute normalization of the measured
differential cross sections was determined from an optical model analysis of
the elastic scattering measured at the more forward angles using potential
parameters found to fit accurately the lower energy data for the same reaction
\cite{Be92}.

A typical two-dimensional charge (Z) versus velocity (v) contour plot of the
fragment invariant cross sections is given in Fig.1 for $\theta$$_{lab}$ = -
7$^{o}$. The velocity v has been deduced by using the empirical mass formula of
Charity and collaborators \cite{Ch88} on the basis of an event-by-event
analysis. Three distinct zones can be clearly identified. The first, where Z is
close to that of the projectile (Z = 17), corresponds to the quasi-elastic
group having a mean velocity close to that of the projectile (v$_{P}$). For
higher Z (Z $\geq$ 18) the fragment velocities are well centered around the CN
recoil velocity (v$_{CN}$) and thus correspond to fusion-evaporation residues
(ER). The third class of events is distributed on two, well separated velocity
branches which merge together as Z is increased, as expected for binary-decay
processes in a reverse kinematics reaction. It is observed that the two
branches which correspond to the two allowed kinematic solutions, overlap
gradually as the detection angles are increased. The two groups belonging to
the second solution arise either from asymmetric-fission components (6 $\leq$ Z
$\leq$ 10) or from symmetric-fission components which are mixed with
deep-inelastic (DI) collisions events (11 $\leq$ Z $\leq$ 15) with large mass
transfer.

In the following, the properties of the binary fragments belonging to this last
class of events will be discussed in detail. These fragments arise from fully
energy-damped reactions. This fact is illustrated by the solid lines drawn in
Fig.1 which have been calculated by using the Viola systematics \cite{Vi85} for
symmetric-fission fragments. The velocities of the non-symmetric mass fragments
have been corrected by the following asymmetric factor \cite{Wi80,Dj92}:
4Z$_{1}$Z$_{2}$/(Z$_{1}$+Z$_{2}$)$^{2}$, where Z$_{1,2}$ are the charges of the
outgoing fragments.

\vskip 2.5 cm

\centerline {\bf III. EXPERIMENTAL RESULTS }

\vskip 1.5 cm

Inclusive kinetic energy spectra were measured for each Z fragment produced in
the $^{35}$Cl+$^{12}$C reaction at E$_{lab}$ = 278 MeV and presented in Fig.2
for $\theta$$_{lab}$ = - 7$^{0}$. The heaviest fragments belonging to the
second zone, discussed in the previous section and displayed in Fig.1, have
typical ER energy spectra arising from the statistical decay of the fully
equilibrated CN formed in a complete fusion process. This is confirmed by the
excellent agreement found for ER with Z $\geq$ 18 with the expectations of the
Monte Carlo code LILITA \cite{Go81} as shown by the black histograms of Fig.2.
It is worth noting that, according to the fusion systematics of Morgenstern and
collaborators \cite{Mg84}, less than 5 $\%$ of the observed ER yield is
expected to arise from an incomplete fusion process. The ER components of the
14 $\leq$ Z $\leq$ 17 energy spectra have been extracted from other
binary-reaction components (yields have been generated by a model to be
discussed in Sect.IV) with the aid of the LILITA simulations at each angle. The
energy spectra of the lightest of the fragments (5 $\leq$ Z $\leq$ 12) are
dominated by the third class of events discussed in the preceding section and
have typical characteristic Gaussian shapes whose centroids correspond to
binary breakup with full energy damping, consistent with the Viola systematics
\cite{Vi85}. The increasing yields at low energy, near the experimental energy
threshold, arise from the second allowed kinematic solution.

The centroids of the first kinematic solution have been extracted for each Z in
order to deduce their total kinetic energy (TKE) values assuming two-body
kinematics in the center-of-mass (c.m.) frame. The results are shown in Fig.3.
The independence of the TKE's and of the differential cross sections
d$\sigma$/d$\theta$ (not shown) on the scattering angle for each exit channel
indicate that the lifetime of the dinuclear complex is longer than the time
needed to fully damp the energy in the relative motion of FF and DI processes.
The average TKE values, also plotted in the insert of Fig.3 as a function of Z,
are found to be very close to the values extracted at lower incident energies
\cite{Be89,Be92}, with only small variations with the incident energy, in
contrast to what can be expected for a DI orbiting mechanism \cite{Ay88}. The
dashed line is the result of a calculation of the equilibrium model for
orbiting \cite{Ay88} with the parameter set used previously at lower bombarding
energies \cite{Be92}. The large discrepancies are essentially due to an
overestimation of the TKE rotational term induced by neglecting diffuse-surface
effects in the calculations performed without corrections for secondary
light-particle emission. On the other hand, the results of FF model predictions
(which will be discussed in the following section), including both these
effects, are shown by the solid line to be very close to the mean values of
experimental data. Furthermore the average TKE value corresponding to symmetric
mass splitting is close to the prediction of Viola \cite{Vi85} and to that of
more recent systematics well suited for light heavy-ions \cite{Be96}.

The experimental elemental Z distribution (full points) of the integrated
fully-damped yields (for Z $\leq$ 12) and ER cross sections are plotted in
Fig.4 with two statistical model calculations (histograms) which will be
discussed in detail in the next section. Because of the potential mixing with
large DI components (which might be composed of either partially or
fully-damped yields), no attempt has been make to extract the FF yields for Z =
13 and 14. The total FF and ER cross sections are $\sigma$$_{FF}$ = 25.0 $\pm$
4.5 mb and $\sigma$$_{ER}$ = 763 $\pm$ 100 mb respectively. The corresponding
critical angular momentum is L$_{crit}$~=~27.5~$\pm$~2.5~$\hbar$ as calculated
by using the sharp cutoff approximation. This value is taken as an input
parameter for the statistical model calculations discussed in the next section.

The possible occurrence of ternary processes that involve three massive
fragments in competition with the binary-decay mechanisms has been searched for
in the fragment-fragment coincidence experiment. The angular correlations
obtained are displayed in Fig.5 for the indicated charge partitions and angle
settings. They are found to peak at well defined angles between $\theta$$_{2}$
= 30$^{o}$ and 50$^{0}$, independent of the charge partition, indicating that
the fragments have a dominant two-body nature as expected for energies lower
than 10 MeV/nucleon \cite{No80,Pe81,Wi81}. The peak positions in the
correlation functions are a measure of the reaction Q-value for the primary
decay. As an example the large and narrow peak for Z$_{1}$=17 and Z$_{2}$=6 has
the position expected for the elastic scattering, whereas the second and
smaller peak at 55$^{0}$ might correspond to a fast neutron transfer process.
Similar narrow peaks are observed at comparable angles in other correlations
with (Z$_{1}$=17 and Z$_{2}$=5) and  (Z$_{1}$=16 and Z$_{2}$=6) that may be
attributed to quasielastic proton stripping or pick-up processes. More
fully-damped processes have larger width correlations which are peaked at
smaller angles. The role of secondary light-particle emission will be to
broaden these distributions, but without significantly affecting the centroids
of the correlations.

In previous experiments using projectiles of mass A$_{proj}$ = 32 to 40 on
various targets, events corresponding to the emission of three heavy fragments
(A $\geq$ 5) have been found to occur significantly (at a 10 $\%$ level) only
at higher bombarding energies (10-15 MeV/nucleon) \cite{Wi84,Pe85,Pe86}. More
recently, however, there has been evidence cited in the literature
\cite{Va88,Bo93} for three-body events in $^{32}$S induced reactions at lower
energies (4-6 MeV/nucleon).

In the following we investigate this possibility of three fragment emission in
the present exclusive data through the analysis of the Z$_{1}$-Z$_{2}$
coincident yields which have been energy integrated for Z$_{1,2}$ $\geq$ 5. The
Z$_{1}$-Z$_{2}$ correlation results are displayed in Fig.6 for the indicated
angle settings. The diagonal lines given by Z$_{1}$+Z$_{2}$ =
Z$_{proj}$+Z$_{target}$ = Z$_{CN}$ = 23 correspond to binary reactions with no
light-charged-particle evaporation. The majority of events are found to occur
near Z$_{tot}$ = 20-21, regardless of whether the exit-mass partition is
symmetric or not. Thus the most probable missing charge $\Delta$Z was found to
be around 2 charge units, which is most likely lost through particle emission
from either the excited composite system or a secondary sequential evaporation
from one of the binary-reaction partners. In order to perform a more
quantitative analysis of these processes we have plotted in Fig.7 and Fig.8 the
coincident yields as a function of the missing charge for the chosen angle
settings. These results are discussed in the next section.

\vskip 2.5 cm

\centerline {\bf IV. DISCUSSION AND CONCLUSION }

\vskip 1.5 cm

The spectra of the summed missing charge, displayed in Fig.7, have typical
Poisson-like distributions with most probable value $\lambda$ = $< \Delta Z>$:
\\

$$\rm P(\Delta Z) \propto \lambda^{\Delta Z} e^{- \lambda} /\Delta Z !$$

The corresponding fits by using this expression are also given in the figure.
The angular dependence of the maxima gives an estimate of the energy
transferred into the fragments according to the two-body kinematics. It should
be noted that a non-statistical emission, such as a three-body break-up
mechanism, will produce enhanced yields superimposed on the exponential
decrease of the Poisson-like shapes at large $\Delta$Z values as shown
previously for $^{32}$S induced reactions at 10 MeV/nucleon bombarding energies
\cite{Be83}. The data of the individual missing charges of Fig.8 are
furthermore reasonably well reproduced by a statistical model calculation that
will be presented later in this section.

An average charge-deficit value of $< \Delta Z>$ = 1.74 is then obtained when
only the fully-damped events are taken into account. This value is appreciably
larger than the one measured for the same reaction at E$_{lab}$ = 200 MeV
\cite{Be92,Dj92} $< \Delta Z>$ = 0.96. This result confirms that the
charge deficit increases linearly with the c.m. bombarding energy and thus with
the total excitation energy available in the composite system \cite{Wi81} and
indicates that the emission process is the statistical evaporation from
equilibrated nuclei. To illustrate this the average charge-deficit values
obtained in the present work along with a collection of other data taken from
the literature \cite{Pe81,Wi81} have been plotted against the c.m.
bombarding energy in Fig.9 as proposed by Winkler et al. \cite{Wi81}. The two
$^{35}$Cl+$^{12}$C data points are shown as stars, whereas the point measured
for the $^{35}$Cl+$^{24}$Mg reaction \cite{Be95,No96} is shown as an
open cross. The linear dependence is fitted by the following relationship :\\
\centerline {$< \Delta Z>$ = 0.048 (E$_{c.m.}$ - 37.12),}
where the c.m. energy is in units of MeV. This behaviour is shown as a straight
line in Fig.9. An energy threshold of about 37 MeV is found for the emission of
light-charged particles and an excitation energy increase of 21 MeV is required
on average for the emission of one unit of particle charge in qualitative
agreement with previous analyses \cite{Wi81,Be83}. Similar conclusions have
also been reached from inclusive measurements of ER mass distributions
\cite{Mo83} and from exclusive measurements on the decay of projectile-like
fragments in the intermediate energy domain \cite{Bea95}. These results suggest
that the emission occurs as a statistical evaporation from equilibrated nuclei.

In summary, the present charge-deficit results are consistent with a
statistical decay of binary fragments and follow the proposed systematics for
this behaviour quite well, in contrast to the data of \cite{Va88,Bo93}. The
absence of ternary events in the present measurement is consistent with results
from $^{32}$S induced reactions where evidence of three-body processes is only
seen at incident energies higher than 10 MeV/nucleon \cite{Be83}. It can be
surmised that, in the present experiment, the inclusive cross sections measured
for the lightest Z fragments (Z $\leq$ 12) arise from a fully-damped binary
process, such as FF, followed by a sequential emission of light-charged
particles and neutrons. In the subsequent discussion we will consider these
fragments as FF fragments.

In Fig.10 the strongest FF channels for the $^{35}$Cl+$^{12}$C reaction
measured at E$_{lab}$ = 280 MeV in this work are presented individually for
each element along with the previously published data \cite{Be89,Be92,Yo89}
between E$_{lab}$ = 150 MeV and 200 MeV. This provides experimental elemental
excitation functions to which statistical-model calculations can be compared.

The FF cross sections rise rapidly with increasing bombarding energies and then
more slowly at higher energies. This behaviour is a characteristic signature
of a statistical CN emission. Therefore it is not surprising that the
experimental elemental excitation functions are very well explained in the
framework of statistical model calculations \cite{Sa91,Ma91} as shown
by the results given in Fig.4 for two types of models.

The first model is based upon the transition-state theory \cite{Sa91} for which
the fission width is assumed to depend on the available phase space of the
saddle point. The second model corresponds essentially to an extension of the
Hauser-Feshbach formalism \cite{Ma91,Ma96} which treats $\gamma$-ray emission,
light-particle (n, p, and $\alpha$) evaporation and FF as the possible decay
channels in a single and equivalent way. The Extended Hauser-Feshbach Method
(EHFM) assumes that the fission probability is proportional to the available
phase space at the scission point. Both calculations start with the CN
formation hypothesis and then follows the system by first chance binary fission
or light-particle emission and subsequent light-particle and/or photon
emission. In the following the full procedure of EHFM, including secondary
emission, will be called EHFM+CASCADE. For instance it is clearly shown from
the EHFM calculations of Fig.8 that the sequential emission plays an important
role in the deexcitation scheme.

In the transition-state model the geometry of the saddle point, including the
role of the fragment deformation, is fully determined by macroscopic energy
calculations. The mass-asymmetric fission barriers are calculated following the
procedure outlined in the liquid drop model of Sierk \cite{Si86} in order to
incorporate effects resulting from finite range of the nuclear interaction and
the diffuseness of the nuclear surface \cite{Kr79}. Both FF and ER yields are
calculated using a modified version of the code CASCADE \cite{Pu77}. The effect
of light-particle emission from the fission fragments (significant at high
excitation energy as shown previously) on the observed element distributions
was simulated using the binary-decay option of LILITA \cite{Go81}. The
transition-state model is an {\it ab initio} calculation based on our current
best understanding of the macroscopic energies of light systems. This
calculation leads to certain results which appear to agree well with experiment
\cite{Sa91}. In the following the full transition-state model calculations with
sequential decay will be labelled as TSM.

The EHFM is an alternative approach using the phase space at the scission point
to determine relative probabilities. In the EHFM calculations the scission
point can be viewed as an ensemble of two, near-touching spheres which are
connected with a neck degree of freedom. The value of the neck length parameter
(or separation distance) {\it s} = 3.0 $\pm$ 0.5 fm is chosen, as is commonly
adopted in the literature \cite{Be92,Na77,Va83,Pu96} for the mass region of
interest. The large value of {\it s} used for the neck length mimics the
finite-range and diffuse-surface effects \cite{Sa91} of importance for the
light-mass systems \cite{Be96,Kr79} and, as a consequence, this makes the
scission configurations closely resemble the saddle configurations of the Sierk
model \cite{Si86}. A systematic study of a large number of systems
\cite{Ma91,Ma96} allows the parameters of the model to be fixed so as to
achieve good agreement with the experimental results. Recent studies
\cite{Ma96} in the framework of EHFM have led to scission configurations being
deduced for the lighter systems being studied \cite{Ma96}.

In EHFM+CASCADE the calculations are performed by assuming first chance fission
which is then followed by a sequential emission of light-charged particles and
neutrons from the fragments. Second chance fission is found, as expected, to be
negligible in this mass region and, therefore, pre-scission emission was not
taken into account in the decay process. The results of the calculated
post-scission emission is illustrated in Fig.8 by a comparison with the
experimental data. EHFM+CASCADE is capable of predicting not only the fission
fragment and ER yields, but also the FF kinetic-energy distributions and their
TKE mean values as shown in Figs. 2 and 3 respectively.

The input parameters of the two models are basically the same. In each case,
the diffuse cut-off approximation has been assumed for the fusion partial-wave
distribution using a diffuseness parameter of $\Delta$ = 1$\hbar$ and
L$_{crit}$ values as calculated from the experimental total fusion cross
sections given in the previous section or taken from previous measurements
\cite{Be89,Be92,Dj92} at the lower bombarding energies.

The predictions of both approaches can be compared to fully-damped yield data
in Figs.4 and 10. A disagreement between the model calculations and the
experimental results as seen in Fig.4 corresponds to a too large observed mass
asymmetry. However, the predictions of EHFM+CASCADE and TSM displayed in Fig.10
provide a quite satisfactory agreement of the general trends of the
$^{35}$Cl+$^{12}$C experimental data over the whole energy range explored. This
might be a good indication of the validity of the hypothesis that the
saddle-point shape almost coincides with the scission point configurations in
this mass region \cite{Sa91,Ma91,Ma96}.

The mass-asymmetric-dependent fission barriers of Sierk \cite{Si86}, which are
central to the success of the TSM calculations \cite{Sa91}, appear to be
appropriate in a first-order analysis of the experimental $^{35}$Cl+$^{12}$C FF
excitation functions. Although more detailed theoretical approaches to the
fission barriers will be needed in this mass region along with other
excitation-function measurements for ``sub-threshold" bombarding energies, the
extracted ``fission thresholds" appear to be quite well understood within a
systematic framework which has been recently established \cite{Be96}.
Experimental studies are being currently undertaken in order to precisely
determine the angular momentum dependence of the mass-asymmetric fission
barriers of light nucleus in this mass region. \\

To summarize, the measured yields of fully energy-damped binary fragments from
the $^{35}$Cl+$^{12}$C reaction at 280 MeV have been analysed as arising from a
fission process, in accordance with previous findings at lower incident
energies \cite{Be89,Be92,Dj92,Ra91,Be93,Sa91}. The coincident data do not show
any evidence for the occurrence of three-body processes, in contrast to recent
observations for a similar system at a comparable energy. The ``charge
deficits" found in the measurement are well described by a complete Extended
Hauser-Feshbach statistical-model calculation which takes into account the
post-scission light-particle evaporation and, thus, can be well understood as
the result of the sequential decay of hot binary fragments. This is in
agreement with the systematic behaviour that has been established for other
reactions studied at bombarding energies below 10 MeV/nucleon. The experimental
$^{35}$Cl+$^{12}$C elemental FF excitation functions have been successfully
described within the framework of the statistical model based on either the
saddle point picture or the scission point picture. The
mass-asymmetric-dependent fission barriers needed in the transition-state model
calculations are found to be appropriate in this mass region.

\vskip 2.5 cm

\centerline {\bf ACKNOWLEDGMENTS }

\vskip 1.5 cm

The authors wish to thank the Post-accelerated Tandem Service at Saclay for the
kind hospitality and the technical support. S. Leotta and S. Reito from Catania
are also warmly thanked for their assistance during the set-up of all of the
experiments. One of us (S.J.S.) would like to acknowledge the NSF for support
within the framework of a CNRS/NSF collaboration program and the U.S.
Department of Energy for support under the Contract Number DE-FG02-89ER40508.

\newpage


%
%
\begin{figure}

Fig.1 : Two dimensional charge versus velocity contour plot of the fragment
invariant cross sections measured for $^{35}$Cl+$^{12}$C at E$_{lab}$ = 278 MeV
at $\theta$$_{lab}$ = - 7$^{o}$. The dotted-dashed line corresponds to the
CN recoil velocity whereas the dashed line is the velocity of the projectile.
The solid lines have been calculated using the Viola systematics as discussed
in the text.
\end{figure}

\begin{figure}
Fig.2 : Experimental (solid lines) inclusive energy spectra measured for
$^{35}$Cl+$^{12}$C at E$_{lab}$ = 278 MeV at $\theta$$_{lab}$ = - 7$^{o}$. The
dashed lines are the results of the EHFM+CASCADE calculations discussed in the
text whereas the black histograms are ER energy distributions as calculated by
the Monte Carlo code LILITA. The results of the calculations have been
arbitrarily normalized to the data for the sake of clarity.
\end{figure}

\begin{figure}
Fig.3 : Laboratory-angle dependence of the center-of-mass TKE values for
the fully-damped fragments from the $^{35}$Cl+$^{12}$C reaction as measured at
E$_{lab}$ = 280 MeV. Averaged TKE values are plotted as a function of the
atomic number in the insert along with DI model (dashed line) and EHFM+CASCADE
(solid line) calculations discussed in the text.
\end{figure}

\begin{figure}
Fig.4 : $^{35}$Cl+$^{12}$C elemental distribution (points) measured at
E$_{lab}$ = 280 MeV compared to two statistical model calculations discussed in
the text. The open and full histograms correspond to EHFM+CASCADE and
TSM calculations respectively.
\end{figure}

\begin{figure}
Fig.5 : $^{35}$Cl+$^{12}$C experimental angular correlations between two heavy
fragments with charges Z$_{1}$ $\geq$ 5 and Z$_{2}$ $\geq$ 5 measured at
E$_{lab}$ = 278 MeV. The first fragment is detected at a fixed angle
$\theta$$_{1}$ = - 7$^{o}$ whereas the second one is detected at a variable
angle $\theta$$_{2}$. The dashed lines are plotted as a guide to the eye.
\end{figure}

\begin{figure}
Fig.6 : Cross sections for coincidence events between two heavy fragments with
charge Z$_{1}$ and Z$_{2}$ measured respectively for $^{35}$Cl+$^{12}$C at
E$_{lab}$ = 278 MeV for the indicated angle settings for which $\theta$$_{1}$ =
- 7$^{o}$. The size of the squares is linearly proportional to the relative
intensity of the pair. The solid lines correspond to binary reactions without
light-charged-particle emission from the fragments.
\end{figure}

\begin{figure}
Fig.7 : Summed charge deficits as measured for $^{35}$Cl+$^{12}$C at E$_{lab}$
= 278 MeV for the indicated angle settings for which $\theta$$_{1}$ = -7$^{o}$.
The solid lines are Poisson distribution fits as explained in the text.
\end{figure}

\begin{figure}
Fig.8 : Individual charge deficits (solid histograms) as measured for
$^{35}$Cl+$^{12}$C at E$_{lab}$ = 278 MeV for each charge with the chosen angle
setting. The dashed histograms are the results of the EHFM+CASCADE
calculations discussed in the text.
\end{figure}

\begin{figure}
Fig.9 : Systematics of the measured charge deficits. The dotted line is the
result of a least-square fit procedure discussed in the text. The stars and
open cross symbols correspond to the data presented in this work, whereas the
other symbols are results taken from other works.
\end{figure}

\begin{figure}
Fig.10 : Experimental $^{35}$Cl+$^{12}$C FF elemental excitation functions
(points) compared with the TSM calculations (solid lines) and EHFM+CASCADE
calculations (dashed lines) respectively.
\end{figure}

%
%

\end{document}